# Celestial Women in Africa

JC Holbrook
jholbrook@uwc.ac.za
jc.holbrook@ed.ac.uk

## ABSTRACT

The indigenous astronomy in Africa and of Africans exhibits many of the same patterns as indigenous astronomy found in other parts of the world such as with agricultural calendars established by observing celestial bodies as well as other natural phenomenon. Africa is the continent with the most countries which can be used as an indicator of its immense cultural diversity, and given such diversity there are unique local aspects to African indigenous astronomy. This chapter answers the question: How do women appear in the indigenous astronomy of Africans? The celestial bodies considered are the Sun, the Moon, the planet Venus, the Pleiades asterism and a small collection of other female celestial bodies. Examples are drawn from North, South, East, West and Central Africa with the caveat that these are merely examples from the region and not exemplars of the region. Women use the phases of the moon to track their menses, usually the moon is female as is Venus, and celestial women are used to signal how women should behave. Focusing on celestial women brings a different lens to cultural astronomy research that elucidates additional ways that that the sky is entwinned in culture.

## 1. Introduction

The continent of Africa hosts over fifty nations and several thousand language groups. The study of cultural astronomy in Africa has been approached in a variety of ways from several disciplinary stances and utilizing their connected disciplinary data collection methods (Starr 1990; Doyle and Frank 1997; Oxby 1999; J. Holbrook, Medupe, and Urama 2008; Alcock 2014; Medupe 2015; Roberts 2015). There are themes connected to the sky found in nearly every African culture such as calendars, navigation, weather prediction, myth, art, architecture and religion as well as commonly named celestial bodies such as the Sun, the Moon and the Milky Way (Blier 1987; J. C. Holbrook 2015; J. Holbrook 2015). Following the lead of previous scholars (Lagercrantz 1952; 1964; Medupe 2015; Roberts 2015), I have opted to structure this African cultural astronomy chapter by focusing on specific celestial bodies; however, I have chosen to focus on the 'female' celestial bodies including goddesses, women and girls. Focusing on these celestial women provides a lens into how the idealized role of women gets projected onto the sky, and how the behaviour of celestial bodies gets projected onto women. Also, focusing on the names of celestial bodies, the cultural function of celestial bodies and the myths associated with celestial bodies reveals some of the ways that cultural astronomy, gendered bodies and gender roles are entangled in the African sky. Finally, given the size of Africa and the number of cultures in Africa, I made the decision to focus on a selection of cultures from north, south, east, west and central Africa neglecting the island nations.

## 1.1. Focusing on Women

For this introduction to African cultural astronomy, I adopt a gender approach inspired by Anthony Aveni's chapter, 'Gendering the Sky,' in his book, *Star Stories: Constellations and People* (Aveni 2019). In that chapter, Aveni explores female constellations and asterisms and the stories related to them. He presents star stories from North America, Asia and Africa. Aveni points out that in the mythology connected to celestial bodies, the women are often portrayed negatively or have bad things happen to them; however, I find that this is not the case for Africa where women often are portrayed as nurturing and creative. With over 3000 language groups and over 50 countries, as I mention previously, the cultural astronomy of Africa is unwieldy. Focusing on women and gender within this context serves both to limit and to focus the materials covered, as well as expand the way that we think about African cultural astronomy and the ideas with which we think. The Cosmic Hunt is related to myths of hunters chasing an animal, but the animal transforming into a constellation – thus it escapes into the sky. This chapter is a cosmic hunt in that I chase elusive sky goddesses such as Tanit of Carthage, spotlight female asterisms such as the Pleiades in many cultures, and assess the role of the planets such as Venus in local cultures in Africa; my hunt is for those women that have become part of the African sky. Particularly difficult was finding the few female Suns among the hundreds of male Suns, and likewise the male Moons among the hundreds of female Moons though these male moons are not given their own section since they are off topic. Rather than chasing women into the sky, I am enticing them to briefly come down to Earth.

## 2. The Women Moon Connection

The physical properties of the moon are often seen as being connected to women's menses, their fertility, and pregnancy, thus many African cultures consider the moon to be female. The Moon goes through phases, that is, it changes appearance because of how it is illuminated by the Sun as seen from the Earth. The new moon occurs when the Sun and Moon appear close enough in the sky such that the moon cannot be seen. The crescent phases of the moon bracket the new moon, with the waning crescent visible in the east near the sunrise occurring before the new moon and the waxing crescent visible in the west after the sunset occurring after the new moon. The term, "waning," means decreasing in size, and "waxing" means increasing in size in this context. Full moon occurs when the Moon appears opposite the Sun in the sky, thus as the Sun is setting in the west, the full moon rises in the east. The female menses of approximately 28 days and the cycle of the moon's phases of approximately 29.5 days are synchronous enough that many cultures connected the two. That relationship between Moon and menses was sometimes given a causal link, such as the Moon being identified as controlling women's fertility and other forms of fertility, and if assigned a gender, the Moon is more likely to be female. The Moon appears to increase in size to its full phase and then decrease to quarter phase and onto crescent phase before disappearing into the sunrise, only to reappear a few days later as a slender crescent just near sunset. Many cultures connected the full phase of the moon as equivalent to a woman being fully pregnant, with the other phases being the weight gaining as the pregnancy progresses and losing weight after giving birth. These physical properties of the moon connect the moon to women, and thus, in most African cultures, the moon is female.

   The Dinka people of Sudan and South Sudan have a female moon as Leinhardt writes (Lienhardt 1987, 200), "For the Dinka the moon is female; for obvious reasons she is connected in thought with women and wives…Women count their periods by the moon." On a practical level, probably the Dinka women note the phase of the moon when their menses starts then they know the next month that when that moon phase approaches their menses will be starting as well.

   In West Africa, there is the Bondo women's society among the Temne of Sierra Leone. Girls learn about marriage, songs, dances, traditional culture as well as probably female related diseases and cures (Berry 1912, 40). During Bondo ceremonies, the Bondo spirit is present in the form of a dancer wearing a Nöwo mask and costume (Lamp 1985). Bondo presides over the girl-to-women initiation ceremony, making a public appearance as the initiates are reintroduced to the community. There is directionality associated with women, spirits, and childbirth where east is the place of origin and can be synonymous with up, thus the Nöwo mask may have symbolic figures on the top indicating the descent of the spirit from the sky through the top of the head of the pregnant woman into the foetus (Lamp 1985, 37, 39 & 41). East being the place of birth and rebirth is associated with the new moon, considered to be up in contrast to the full moon considered to be down, this points to heliacal observations just before sunrise, that is, the Sun rises in the East as the full moon is setting in the west. The waxing crescent which is the beginning of their month is also associated with east; the crescent moon either waxing or waning is depicted on some Nöwo masks and other objects.

   The Fon people of West Africa, found in Benin, Togo, and Nigeria, have Mawu-Lisa, a sky deity that is both male and female as reflected in the Sun and Moon (Lawal 2008, 27). The female aspect is Mawu, and connected to the moon, she rules the night and fertility among other aspects (Melville Jean Herskovits and Herskovits 1998, 125). Additionally, Mawu is a creator in that she makes the first humans and recycles their dead bodies to make new humans (Ikenga-Metuh 1982, 16). Herskovits (1932) provides a description of the Mawu-Lisa Sky-God cult and rituals, which he puts forth as more opulent than those of the other religious cults. Mawu-Lisa has a divided connection to the daily motions of the sun in that Mawu (female aspect) is connected to the rising sun, while Lisa (male aspect) is connected to the setting sun (Olupona and Kunnie 2015, 8).

   In Southern Africa, among the Pedi of South Africa, the moon and its phases have layers of meaning, some connected to women and others to life more generally (Vogel 1985). The waxing crescent moon and waning crescent moon appear in wall murals symbolizing the beginning of a woman's life and the end of her life, but also the beginning of her fertility cycle and the end, with the full moon being her most fertile time (Vogel 1985, 82). The waxing and waning crescents are symbolic of life's beginning, adulthood, and end as well as of agricultural cycles.

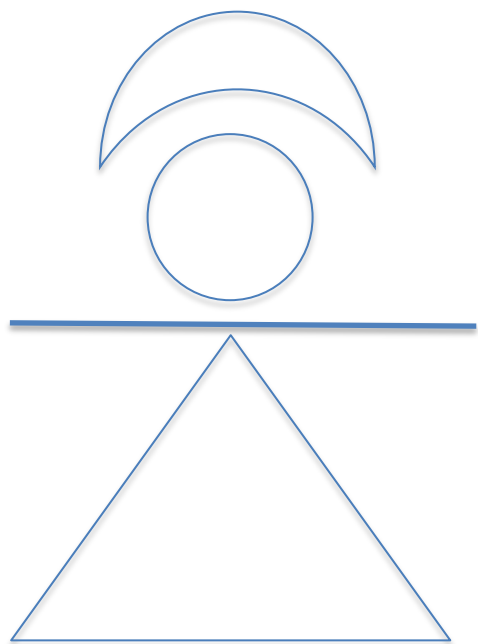

Figure 1: Sign of Tanit. A symbol including the solar disk and crescent moon.

For many centuries the goddess of Carthage, Tunisia, was Tanit (also Tanith, Thanit, Tinit), a moon goddess of fertility. Her symbol is recognizable as a triangle, line, solar disk, and crescent moon and found on artefacts from the region (Moscati 1968, 139). Piecing together snippets of information about Tanit from commentary about works of art, Tanit became the patron deity of Carthage from about the 5th century BCE (Clifford 1990, 61), Brody (Brody 1998) makes the case that Tanit is a protector of seafarers; her consort is Ba'al Hammon, who was a solar deity, though she was placed above him in the hierarchy, and the Sign of Tanit incorporates both deities in that she is sometimes indicated as the Face of Ba'al (Nierfeld 2008, 581). The legend is that Tanit was invited through a spoken ritual, "*evocatio*," to relocate to Rome after the fall of Carthage in 146 BCE; however, whether that *evocatio* actually occurred has been debated (Le Gall 1976; Greene 1996; Isaenko 2019). Tanit was Romanised into the goddess Caelestis (Varner 1990, 12; Hvidberg-Hansen 1986).

In East Africa, the Arimi of Tanzania regard the moon as male or female depending upon its phase (Jellicoe, Puja, and Sombi 1967). When the moon is a waning crescent, it is considered male; but as the moon waxes to full, it is considered female: "As the Moon rises to the full she becomes female, pregnant with all living things; when she disappears from the sky; children and animals are born and women fall into menstruation. The most auspicious time of the month is when the full Moon appears in the east as the Sun sets in the west; at this time all major annual ceremonies should ideally reach their climax" (Jellicoe, Puja, and Sombi 1967, 29).

To the East of the Arimi are the Sandawe people of Tanzania. Their cosmology shares some overlap with the Arimi linguistically, but their moon is female independent of phase. In their creation myth, the Moon is the first celestial body, and during that time, the Earth was perceived as a paradise because it was both cool with plenty of rain and fertile (Raa 1969). The supernatural being that would become the Sun fell in love with her and joined her in the sky drying out the Earth. The Moon had to make rain to return fertility to the Earth, and upon completion, the Sun and Moon's children, humans, were released upon the Earth. As with the Arimi, among the Sandawe, the Moon is connected to and is the source of fertility and with the menstrual cycle of women. Further, the Moon is thought to determine the sex of the child depending upon its phase during conception: waxing phases will produce females, whereas the full moon and waning phases will produce males (Raa 1969, 30). The layered meaning here is that "infancy and growth are associated with femininity, and fullness and decline with masculinity" (Raa 1969, 31).

The G/wikhwena San people of Botswana, similar to the Arimi of Tanzania, also have a moon that changes gender depending upon its phase (Silberbauer 1981, 126). When the moon is waxing (increasing in size) it is considered male; whereas when it is waning (decreasing in size), it is considered female. This is the opposite of the Arimi, where waxing is female and the waning is male.

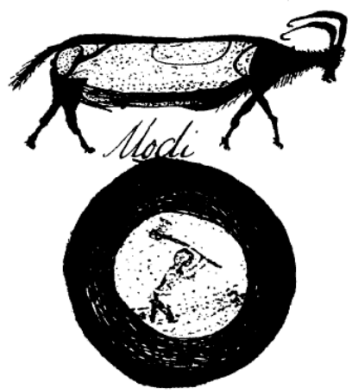

**Figure 2: Detail from illustration in Keller & Huber (1903). The circle is the moon where the woman wielding her axe is shown. DO NOT REPRODUCE RIGHTS ARE BEING NEGOTIATED.**

The Subu people of Cameroon have a legend of the dark features of the moon that they consider to be a woman (Mesendi m 'ewondo). The woman gathered firewood and chopped wood on their sacred day of rest and was punished by God. She was put up in the moon where she can be seen with her axe (Keller and Huber 1903). Thus, her location in the moon is punishment for breaking the rule of not doing work on their day of rest.

The moon tends to be female among Africans because its phases are similar to the weight gain and loss associated with pregnancy as found in the moons of the Arimi people, because its phases have a cycle of similar length to the menses cycle of women as with the Dinka people, and thus the moon is connected to fertility more generally as with the Arimi, Fon and Sandawe people as well as the Carthaginians. The Moon is the first celestial body examined in this chapter where a celestial body changes gender, which is the case among the G/wikhwena San and the Arimi people. I return to gender changing later when discussing the Sun.

### 3.  Pleiades as Star Women

The tight cluster of stars the Pleiades are recognizable riding the back of Taurus the bull in the standard Western constellations. Most people see six stars in the cluster but telescope images show many more stars that are too faint for human eyes to see. Many legends have the stars of the Pleiades as women, with an occasional story of them as groups of men or boys, and at other times as a clutch of eggs (Aveni 2019). Layers of meaning often include the appearance and disappearance of the Pleiades from the night sky as a calendar marking seasonal variations such as wet and dry season, harvest or planting time, or the time for hunting certain animals (Frazer 1913; Hirschberg 1929; Snedegar 1997). For example, the heliacal rise of the Pleiades is a marker for the beginning of the New Year for many African cultures. The heliacal rise of a star or constellation refers to it becoming visible in the night sky just before sunrise, which means the return of that celestial body to the night sky for many months until it disappears into the sunset. Not all celestial bodies rise and set, there is a dependence upon the latitude location of the person doing the observations. In the tropics, all the stars rise and set; in the temperate regions, some stars rise and set while others circle the celestial pole; in polar regions, most of the stars circle the pole without rising or setting. Africa straddles the equator and includes the tropics and temperate regions, thus their night sky may include heliacal rises. Bennet (2018) summarizes that the Zulu started their new year with the heliacal rise of the Pleiades, whereas additionally the Zulu, the KhoiKhoi, the Setswana (Batswana) associated the position of the Pleiades in the night sky, for example when the Pleiades is overhead at sunset, with indicating the time to start cultivation. Among the people of Zambia, the Lozi and the Ila use the heliacal rise of the Pleiades to start their new year, the Senga use the Pleiades rising in the east at sunset to start their new year; additionally the Nyanya, the Lozi, and the Ila use the position of the Pleiades at sunset to start their cultivation (Chaplin 1967).

The Pleiades name is associated with the Greek myth of Orion and the Pleiades. The myth has Orion chasing the seven sisters and capturing one, Merope. Thus, there are six stars in the Pleiades with the seventh missing. However, when the Pleiades are near the western horizon at sunset, Venus can be close enough to appear to be the seventh Pleiad, and night after night, Venus appears to move in the direction of Orion. It is as if she has been captured by Orion (Figure 3). This occasional celestial event may be the source of the Orion legend and may be why the Pleiades are often referred to as a set of seven rather than of six.

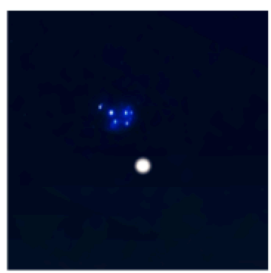   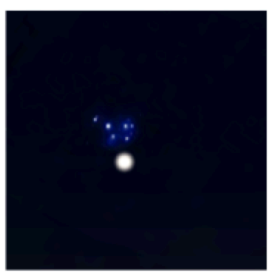   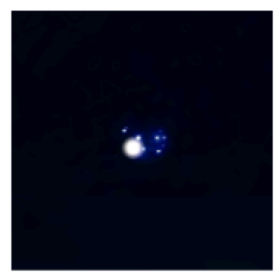

April 1, 2020                        April 2, 2020                        April 3, 2020

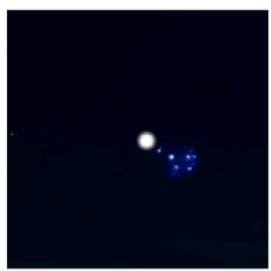   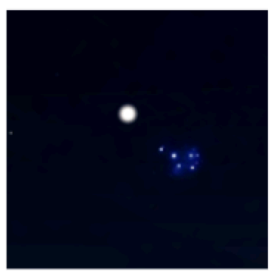

April 4, 2020                        April 5, 2020

**Figure 3: A Stellarium simulation of Venus joining the Pleiades briefly in April 2020 – Stellarium Software is used for simulating the night sky (Zotti and Wolf 2018). Orion does not appear in these images, he lies far to the left of these images. (Stellarium 0.20.1 - https://stellarium.org/ Copyright (C) 2000-2020 Fabien Chereau et al. & the Pleiades pictures come from Herm Perez: http://home.att.net/~hermperez/default.htm)**

Among the Tuareg people of the Sahara, the Pleiades are 'The Girls of the Night' where six have names (Duveyrier 1864, 425): Mâteredjrê, Erredjeàot, Mâteseksek, Essekâot, Màtelarhlarh and Ellerhâot. The seventh is the eye of a boy that left his head and went into the sky. The lack of a proper name for the seventh star and the difference of the seventh star in that it is both male and an eye might reflect the seventh being the occasional visitor Venus, as with the Orion myth. However, I have been unable to find more information that would provide confirmation.

In Tanzania, the Arimi people use the return of the Pleiades to the sky in September as an indication of the end of the dry season (Jellicoe, Puja, and Sombi 1967). Named 'Kiimia', "she is supremely the wife and mother, connected with green and yellow, the colours of grass and flowers. In September she rises from a long dry-season embrace with the great python in the rain shrines in order to co-operate with the Sun and Moon to bring to fruition that which they have produced between them" (Jellicoe, Puja, and Sombi 1967, 29). The ideal characteristics of local women are attributed to the Pleiades such as taking care of babies and baby animals and offering water to strangers in need. "Kiimia also protects people from the anger of Murungu by closing the hole made in the sky by the lightning" (Jellicoe, Puja, and Sombi 1967, 30). The Pleiades as conceptualized as one celestial body, but the multiplicity of it being a star cluster is alluded to when mentioned as plugging a hole in the sky in the quote. Also, 'Murungu' is the Arimi male sky god. There are additional layers of meanings and symbols connected to the Pleiades, including a cartographical association of seven stars representing the four cardinal directions, center, up and down, drawn as a seven-pointed star. The navel stone called 'Nkhoma' in a rainmaking shrine described in Jellicoe (1967), is associated with the Pleiades. The position of the navel stone is the navel of the rain shrine whose structure represents the body of a woman laid out on an east-west axis.

In Southern Africa, Alcock (2014) recounts a legend among the KhoiKhoi people as documented over a century ago regarding the Pleiades and their husband (Schultze 1907). In the legend, the Pleiades are a group of women in a polygamous marriage with one husband (the bright red star Aldebaran in Taurus). They sent their husband out to hunt for meat but warned him to not return empty handed. Unfortunately, he shot his arrow at a group of zebras and missed. Of course, one has to wonder why he went hunting with only one arrow, but clearly it was to advance the story and connect it to the sky since the arrow is identified as a line connecting three faint stars in Orion: iota, theta, and 42 Orionis (Alcock 2014, 288). A lion (the red star Betelgeuse in Orion) was hunting the same herd of zebra from the other side, near where the arrow fell. The herd of zebra are the belt of Orion: Alnitak, Alnilam, and Mintaka. Thus, the husband remains in the sky unable to return home, whereas the women were boastful, saying to the other men: "Ye men, do you think that you can compare yourselves to us, and be our equals? There now, we defy our own husband to come home because he has not killed game" (Alcock 2014, 287). Wessels explains: "Success in the hunt is critical to gender harmony. Women would not consider staying with a man who did not bring meat back from the hunt" (Wessels 2007, 315). The Khoikhoi terms for these are: ǂab – arrow, aob – the husband, Aldebaran, !goregu – zebras, the three stars in Orion's Belt, /Khunuseti – to gather together, the Pleiades, the wives of the tale (Alcock 2014, 287 & 288).

In the Democratic Republic of Congo, the Bangala people call the Pleiades 'lingondo nsamba', which means a 'crowd of young women' (Weeks 1909a, 417).

Though not women, the Pleiades are considered a female chicken that is a hen with her chicks in many West African cultures. Though not identified as the Pleiades,

Udoh gives 'unen eka ndito' as designating a constellation meaning "the mother hen and the chicks" among the Ibibio and Annang people of Nigeria (Udoh 1977, 11). In the same way, the Hausa dictionary never names the Pleiades but mentions the hen and chickens constellation that appears in the sky when the rainy season is coming, 'kaza mai-yaya' or 'kaza da yaya' (Robinson 1913, 113). Urama (2008, 233) mentions directly that the Pleiades is the Hausa constellation of the hen with chickens (*kaza Maiyaya)* in Nigeria. In Mali the Pleiades are the hen and chicks 'niugu-niugu' (Oxby 1999, 57). The Bulsa of Ghana have 'chibiisa', which is the hen and chickens constellation, the Pleiades (Apentiik 2003, 243).

The Pleiades are female among many African cultures, they are connected with seasonal rains (serving as a calendar marker) and thus fertility. Among many West African peoples, the Pleiades are a hen (female) and her chicks, thus there is a mothering/family role similar to how the Arimi think of the Pleiades as wife and mother. Uniquely, the Tuareg have a boy's eye as the seventh star, whereas the other six stars have female names, which may indicate that the seventh star is Venus, which occasionally but regularly appears close to the Pleiades.

### 4.  Female Suns in Africa?

Africa is dominated by a male Sun, from the Egyptian Sun god Ra/Re to Kuiye the Sun god of the Batammaliba people of West Africa (Blier 1987) to the old man of the |xam people of Southern Africa whose armpits gave off light. Female Suns in Africa are very rare and come in two forms: The Sun is female all the time or the Sun is female some of the time. The latter points to the Sun changing sex depending upon the situation such as whether it is rising or directly overhead and whether the associated temperature is hot or cool.

The Sandawe in Tanzania have a female name for the Sun, '//'akásu', but all of their celestial bodies are given female names because of their smallness (Raa 1969). Regardless of this, the Sun is male, and all its actions are associated with masculinity. The Sun's masculinity is less during the early morning when it is considered mild but becomes more destructive, and equivalently more masculine, as it rises towards the zenith.

Among the G/wikhwena San people of Botswana, the Sun, '/amsa', is female (Silberbauer 1981, 126). This designation is in keeping with their cosmology where things that are hot or cold (among other criteria) are considered female. Similar to the Sandawe, the Sun at the zenith is considered harsh; but there is not a record of it changing gender as it changes position on the Sky among the G/wikhwena.

In Palmer (1914, 116), there is the tantalizing note that "The "moon" is in Hausa and several Berber dialects masculine, the story being that it is a boy ('yaro=eiuro') which the sun ('rava') chases round the sky." The Sun is the mother while the moon is her son. The Hausa dictionary confirms that the Sun is the female noun, 'rana' (Robinson 1913, 247), whereas the moon is not given a gender and 'yaro' means boy. Chasing down references to Berber names for the sun and moon, the moon is male and called 'Aiour' or 'Aggour' according to Basset (1910, 305), but Basset does not mention a female Sun rather provides male deities connected to the Sun. However, chasing down references on the Hausa and the Berber yielded no further information to add to that note.

Consideration of the Sun being female among some African cultures has a positive answer with caveats. The Sun is sometimes female, but in the cases I uncovered, the Sun changes gender depending upon its location; thus it isn't always female. In the Hausa case where the female Sun is a mother chasing the Moon/her son, the Sun is female, but I could not find anything more about the myth. The Sun is female because of linguistic conventions such as everything in the sky being designated female for the Sandawe people and hot things being female for the G/wikhwena San people.

## 5. The Planet Venus

Venus is the planet closest to the Earth with a closest distance of 38 million kilometres away. It is the second planet from the Sun (remember the Earth is the third), and it has a similar but smaller size and mass as the Earth. Venus is bright enough to be seen during the daytime without the aid of a telescope or binoculars if you know where to look. At night Venus is the brightest object in the natural sky besides the moon. Like the moon, Venus appears to go through phases where it increases in size and brightness and decreases in size and brightness depending upon the angle between the Earth, Venus, and the Sun. At maximum magnitude of -4.89 and a minimum magnitude of -3.83 it is easily brighter than the brightest star in the night sky, Sirius, with an apparent magnitude of -1.47.

Venus is prominently visible during the twilight times as the Sun is setting or just before sunrise. It is a dramatic celestial body given its brightness. That brightness is possibly the reason why many cultures consider it both beautiful and female. In Roman mythology, Venus was a goddess known for her beauty, and she ruled over love and fertility among other things. Being beautiful is enough to make Venus female as a celestial body, but there is another connection to women and their cycles. Venus is known as the Morning Star and the Evening Star. The apparent motion of Venus relative to the Sun takes 584 days to complete a full cycle. During each cycle, Venus has a Morning Star phase, an Evening Star phase, and two phases where Venus is not visible. The Evening Star phase is when Venus follows the Sun and sets after the Sun sets, which lasts for 263 days. Venus is then not visible for 8 days before reappearing as the Morning Star. The Morning Star phase is when Venus rises in the east before sunrise, this lasts for 263 days. Finally, Venus is not visible for 50 days before emerging as the Evening Star again. In his book, *Conversing with Planets*, Aveni mentions that human gestation is between 255 and 266 days (Aveni 1992, 102). A synchronicity or resonance exists between the nine months of human pregnancy and the nine months of Venus being visible in the evening sky or morning sky. Knowing this nine-month cycle makes the association of the planet Venus and the goddess Venus with fertility conceivable.

Frobenius (1937, 215) gives us the story of Venus as morning star and evening star and their relation to the moon among the Wahungwe/Hungwe people of Zimbabwe. In the story, the Moon is male, and his first wife is the morning star (called 'Massassi') that he is married to for two years; his second wife is 'Morongo' the evening star. The second marriage was limited to two years as well, but through both marriages, the Moon and his wives gave birth to all the plants, animals and humans.

There are many African cultures where Venus is known as the wife of the moon. In the Democratic Republic of Congo among the people living around the town of Matadi and west to the Atlantic Ocean, 'Nkaz' a Ngonde' is their name for Venus, which translates to wife of the moon (Weeks 1909b, 477). The Tabwe people also of the Democratic Republic of Congo have both aspects of Venus, morning star and evening star, as the wives of the Moon (male) in a polygamous marriage (Roberts 2015, 1039). The Zande people jokingly say that the evening star is the moon's wife called 'Tungu' (Larken 1926, 43); however in their cosmology, everything supernatural including things in the sky are considered gender neutral.

Weeks gives three names for Venus among the Bangala people of the Democratic Republic of Congo (Weeks 1909a, 417): Venus is 'mwali wa sanji', the wife of the moon; the morning star is 'mokwete' and 'yombi'; while the evening star is 'mosolampande' (Weeks 1909a, 418). This may reflect the fact that the morning star and the evening star are not always Venus. Other planets (Mercury, Mars, Jupiter and Saturn) can be seen as the bright first star of the night or the bright last star before morning. These visible planets are bright and can be seen during the twilight – times usually dominated by Venus – though Venus is still the brightest. The planets beyond the Earth – Mars, Jupiter and Saturn - can be seen further from the Sun, whereas Venus and Mercury appear close to the Sun. The Hausa dictionary lists Venus as 'matan wata', which is wife of the moon (Robinson 1913, 274), but also has a different name for the morning star, which is 'gamzaki' (Robinson 1913, 240).

Finally, as mentioned in the paragraphs on the Pleiades, Venus might be the missing seventh Pleiad, and the eye of the young man, which is the seventh star among the Tuareg, may actually be Venus. If correct, then it would be an example of a male Venus in Africa; however, I have been unable to find supporting evidence.

For completeness, Iwaniszewski's analysis of Venus beliefs found that the global trend was for Venus to be male during its morning star aspect and female during its evening star aspect, reflecting the cosmological directionality of East connected to male characteristics and West connected to female characteristics (Iwaniszewski 1996). Such a distinction was not obvious among this selection of African cultures presented, supporting Iwaniszewski's conclusion that African cultures tend to assign a single sex to Venus.

Venus as morning star and as evening star tends to be female in Africa, even among the Zande, who, even though they consider everything in the sky to be genderless, say Venus is the wife of the Moon. Thus, Venus can be both a celestial wife as among many African peoples and mother as among the Wahungwe/Hungwe people.

## 6.  Other Female Celestial Bodies

In addition to the Moon, the Sun, the Pleiades and Venus, there are other celestial bodies that are considered female in Africa; however, which celestial bodies are culturally specific with little to no overlap with other African cultures. The uniqueness of these celestial women/females exposes aspects of the local environment.

In Tanzania, the Arimi people regard the Small Magellanic Cloud as female while the Large Magellanic Cloud is male. During the rainy season, the Magellanic Clouds are thought to aid the Pleiades in bringing the heaviest rains when they appear beside each other (Jellicoe, Puja, and Sombi 1967, 30). The authors do not indicate what time of night to observe the Large and Small Magellanic Clouds; nonetheless, **Error! Reference source not found.** shows a Stellarium simulation of how they would appear to be next to each other in the night sky facing South in Tanzania.

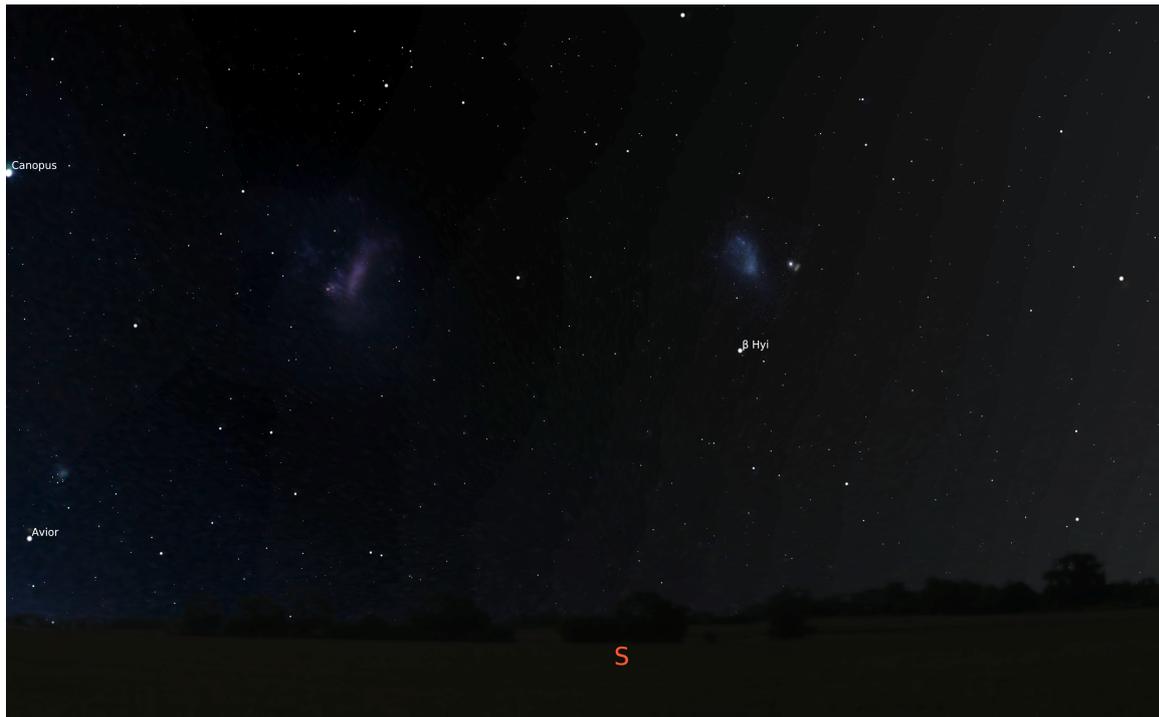

**Figure 4: Stellarium simulation showing the Large (LMC) and Small Magellanic Clouds (SMC). This Stellarium image (Chéreau 2003) uses as location the town of Singida, Tanzania, which is the main city of the region where the Arimi people live. This would be looking South near sunset on January 11th and shows the LMC on the left and the SMC on the right. (Stellarium 0.20.1 - https://stellarium.org/ Copyright (C) 2000-2020 Fabien Chereau et al. & images of SMC & LMC (Magellanic Clouds) from Albert Van Donkelaar.)**

There are the female giraffes ('n//abedzi') represented by the Southern Cross (Crux); and Peacock, the brightest star in the constellation Pavo, is a female steenbok ('g≠eisa') among the G/wikhwena San people of Botswana (Silberbauer 1981, 109).

Though the Milky Way itself is not female, among the |Xam people of South Africa it is a woman that creates the Milky Way (Bleek and Lloyd 2007; Wessels 2007). Wessel provides detailed context to the story. The girl was experiencing her first menses and thus was secluded; she was angry because her mother did not provide her with enough of a special root to eat; she threw ashes and roots into the sky making the Milky Way (Wessels 2007). The girl was special because she had the ability to create, but more broadly supernatural powers are attributed to girls/women while menstruating in several |xam myths. Of interest are other instances when constellations are made: "A group of people who are sitting eating a rock-rabbit are turned into the constellation Corona Australis when a menstruating girl looks at them" (Wessels 2007, 311). The positive aspect of the story of the Milky Way is that through making the Milky Way and other stars, they become navigational aids to guide the men out hunting at night home.

Among the Tuareg people of the Sahara, there is a version of Ursa Major and Ursa Minor where Polaris, the North Star, is a Black Woman (d'Huy 2013, 94). The mother camel is Ursa Major, and the baby camel is Ursa Minor. The Black Woman is afraid the other stars nearby are trying to kill her, so she keeps still. In another version, the stars nearby want to capture her to enslave her. This legend appears in Duveyrier (1864, 424):

> *L'Étoile Polaire est dite Lemkechen, mot à mot, tiens, c'est-à-dire qu'une Négresse est supposée recevoir l'ordre de tenir le Chamillon Aourâ, pour qu'on puisse traire sa mère, Tâlemt, la Chamelle (c'est- à-dire la grande Ourse). Les étoiles de la même constellation ψ, λ, μ, ν, ξ, qui forment un triangle, figureraient une Assemblée, El-Djema'at, qui délibérerait pour tuer Lemkechen (la Négresse); c'est pourquoi cette dernière, saisie d'effroi, ne bouge pas et cherche à se cacher.*

Translated: The Polar Star is called Lemkechen, word for word, hold, that is to say that a Negress is supposed to receive the order to hold the [baby] camel Aourâ, so that one can milk her mother, Tâlemt, the camel (ie the Big Dipper). The stars of the same constellation *ψ, λ, μ, ν, ξ*, which form a triangle, would represent an Assembly, El-Djema'at, which would deliberate to kill Lemkechen (the Negress); this is why the latter, seized with dread, does not move and seeks to hide.

  The stillness of the Black woman reflects the physical reality that Polaris is close enough to the North Celestial Pole that it appears not to move, while the other constellations circle around it. People mistakenly think that Polaris is the brightest star in the sky, but it is moderately bright at 1.97 magnitudes. The fact that it appears to be stationary is what makes it unique. It marks the direction of north and can be used as a navigation marker as long as it is visible. The Tuareg story about Polaris/Lemkechen is the only mention I could find of this important star being female in Africa. Related is 'taguwa' the camel constellation found in the Hausa Dictionary (Robinson 1913, 41).

  Nut is the sky goddess of ancient Egypt (Roth 2000, 195). Her body is the sky, and she swallows the Sun (Re) every night and gives birth to him in the east every morning. The red sky at dawn is her blood that is part of the birthing process (DeYoung 2000, 477). Nut as mother goddess gave birth to the goddess Isis (DeYoung 2000, 475). Isis is the bright star Sirius in Canis Major (Krupp 2000, 4). Isis is married to her brother Osiris represented by the constellation Orion. Osiris is killed by their brother, and his body parts are spread all over Egypt. Isis works to put Osiris back together, and thus Isis follows Osiris across the sky (J. Holbrook 2011) just as Sirius follows Orion across the sky (see Figure 5). Before Isis became the dominant mother goddess in Egypt, Hathor, another goddess of fertility, was associated with Sirius (Teeter 2017, 156). Hathor either has a cow head with a Sun disk between her horns or is depicted as a cow with the star Sirius above her horns (Holberg 2007, 8).

  Finally, I want to note that in Africa there are ancestral women who came down from the sky, such as among the Akan of Ghana (Wilks 2004, 32). The phenomenon of having sky ancestors is a strategy to establish political authority. As Wilks discusses (Wilks 2004, 18), to have ancestors that come from somewhere else means that they immigrated; whereas if they come from the sky or the ground (up or down), they can be considered of the land that they now occupy. These are celestial women that have come down to Earth, and thus ground their people to the land.

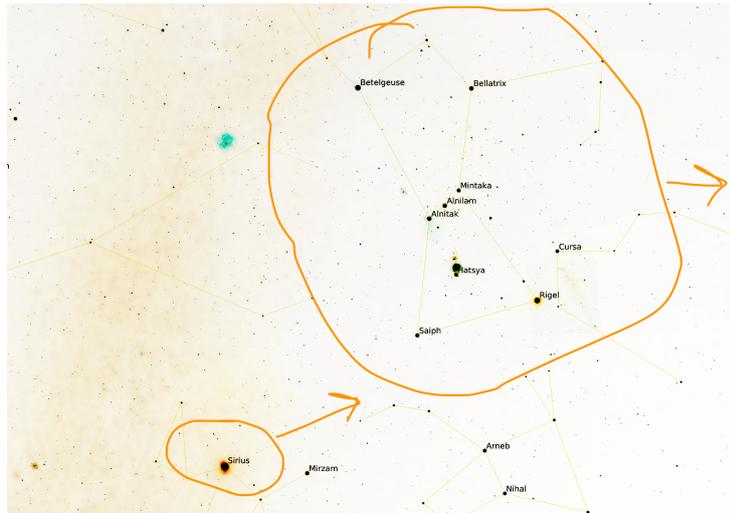

Figure 5: Stellarium Image of Sirius and Orion with arrows showing the direction of motion over the night. (Stellarium 0.20.1 - https://stellarium.org/ Copyright (C) 2000-2020 Fabien Chereau et al.)

Other celestial women and female animals present information about the local environment such as when the rainy season starts among the Arimi people, that the stars can be used for navigation among the |xam and show local animals such camels, giraffes, steenboks and cows. Myths that tie several constellations together and in some ways serve as a mnemonic for remembering the locations of particular constellations and the way that they move across the sky, similar to the Orion/Pleiades myth, are present here in the Egyptian myth of Osiris and Isis (which are Orion and the Pleiades as well) and the Tuareg Lemkechen myth where Lemkechen stand still while the baby camel circles around her. Celestial women coming to Earth are used to establish local authority as among the Akan people. These other celestial women add to the picture of the African gendered sky and Earthly gendered relationships.

## 7. Conclusions:

Africa has a long history of sky watching and a rich collection of skylore that can be used to gain an understanding of those celestial bodies that are culturally important and, in some cases, why they are culturally important. The exercise of framing this exploration using a gendered focus on celestial women yielded expected results that align with celestial women outside of Africa, but also yielded unexpected results such as female suns and male moons. Meaning in many ways Africa is normal, taking its place with the rest of the world in terms of gendered celestial bodies, and also unique with having some atypical gendered celestial bodies such as female Suns and gender changing celestial bodies. In this chapter, I have only touched upon the connection between female celestial bodies and the timing of rituals connected to female initiations as found in West Africa; there are more. Much more common is how African women and others use the phases of the moon to keep track of the menses and therefore their fertility. The Pleiades draws the eye because it is a tight cluster of stars that is bright and considered beautiful and therefore normally is considered female or a group of females. In contrast to the Pleiades, the nearby Hyades cluster rarely is remarked upon except for the mention of its bright red star Aldebaran. Similar to the mythological complexes found in the Greek and Roman skies, likewise there are mythological complexes found in Africa where myths weave stories connecting celestial bodies such as the women (Pleiades) sending their husband (Aldebaran) out to hunt game (Orion's Belt) found among the KhoiKhoi people in Southern Africa. The women found in the sky in Africa are sometimes powerful and magical such as the girl who created the Milky Way among the |xam people, but there are also negative stories such as the women trapped in the moon wielding her axe and chopping wood for all time among the Subu people. Celestial women can be considered role models of idealized female behaviour within particular cultures. For example, there are repeated projections of the gender role of wife and mother placed upon the Moon, Venus and the Pleiades. Relatedly, women are meant to follow the cultural rules placed upon them with sometimes dire consequences such as the dangerous gaze of a young woman experiencing her first menses causing people to be transformed into stars or being trapped in the moon because of working on a sacred day of rest. Women role models, powerful women, reminders to follow rules, regional female animals are all part of the gendered African sky. As with any study attempting to synthesize information from a large geographical region, the work presented here cannot be generalized to represent all of Africa and/or all Africans, but rather it is a snippet that can be the foundation for a more expansive study of celestial women in African skies.